# Mechanical behavior of HTS tape in highly flexible REBCO cable under tensile and torsional loads

Shengyi Tang[1,2,3], Huadong Yong[1,2,3], Youhe Zhou[1,2,3]

1 Key Laboratory of Mechanics on Disaster and Environment in Western China, Ministry of Education of China, Lanzhou University, Lanzhou, Gansu 730000, People's Republic of China

2 Institute of Superconductor Mechanics, Lanzhou University, Lanzhou, Gansu 730000, People's Republic of China

3 Department of Mechanics and Engineering Sciences, College of Civil Engineering and Mechanics, Lanzhou University, Lanzhou, Gansu 730000, People's Republic of China

**Abstract:** One kind of highly flexible REBCO cable (HFRC) has been proposed recent years, which contains REBCO superconducting tapes wound onto a spiral tube. Under external loads, spiral tube is prone to complex deformation, leading to unpredictable characteristics of HFRC. In this paper, we analysis the deformation process of spiral tube under tension and torsion. Based on the geometrical relationship between tapes and spiral tube, the mechanical behavior of tapes in HFRC is further studied.

**Keywords:** superconducting cable, high temperature superconducting tape REBCO, tension, torsion, spiral structure

## 1. Introduction

Advances in superconducting technology have provided a basis for the development of various industries. From nuclear-magnetic resonance spectrometer to nuclear fusion reactor, high temperature superconducting (HTS) materials, such as REBCO ((RE)Ba$_2$Cu$_3$O$_x$) tape, play an important role in the construction of these devices[1-3]. In practice, REBCO tapes need to be assembled into different kinds of power cables, such as TSTC (Twisted Stacked-Tape Cable)[4], TSSC (Twisted-Stack Slotted-Core)[5], RACC (ROEBEL Assembled Coated Conductor)[6], Q-IS (quasi-isotropic) strands[7, 8], and CORC (Conductor on round core)[9]. Given the favorable properties of CORC cable, it has attracted the attention of researchers in recent years[10-19]. This kind of cable consists of two components: one round core and several layers of REBCO tapes, which are spirally wound around the core with opposite helical direction in adjacent layers. The advantages of CORC lies in two areas. From the perspective of mechanics, the manufacturing process is simple, and the round cross-section makes it easy to be assembled into different shapes. From the perspective of electromagnetism, due to the spiral configuration of tapes, its AC loss is lower and the critical current is independent of magnetic field angle. In order to enhance the performance of CORC, researchers have been optimizing its structure[20-22]. For example, the Institute of Plasma Physics, Chinese Academy of Sciences (ASIPP), has proposed a new kind of CORC named as HFRC (highly flexible REBCO cable)[20]. In HFRC cable, the core is replaced from a round rod to a spiral tube, while the tapes of 1st and 2nd layer (closest to the core) are copper tapes (Fig. 1).

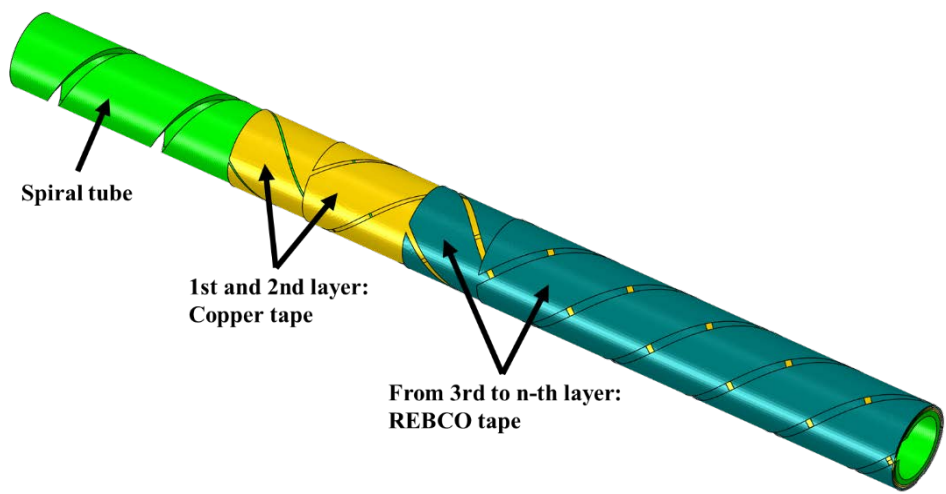

Fig. 1 Configuration of HFRC cable

From Fig. 1, we can find that the significant feature of HFRC is high flexibility, so it can be mounted in magnets with small mechanical loads. Additionally, the spiral tube of HFRC makes room for cooling medium to flow inside the cable[20] and exchange heat with the tapes. Many researchers have tested the performance of HFRC[20, 23, 24]. Guo et al. tested the contact resistance in HFRC, noting that the difference in thermal expansion rate between tapes and spiral tube may enhance the contact resistance[24]. Jin et al. tested the degradation of critical current during the bending process of HFRC, finding that HFRC has a smaller critical bending radius compared to traditional CORC[20]. Xiao et al. studied the behaviors of HFRC during charging. The results show that the performance of tapes after electromagnetic cycles is influenced by the mechanical strength of spiral tube[25]. Shi et al. compare the critical transverse compression loads of CORC and HFRC[26]. Although the above researches have demonstrated the effects of spiral tube (core of HFRC) on the behavior of the REBCO tape under different loads, the properties of HFRC under tensile and torsional loads have been less studied. Therefore, some efforts need to be made in this area.

The behavior of traditional CORC with solid core has been deeply studied[12-15, 26-31], including the processes of tension and torsion. Assuming the core remains cylindrical and the centerline of tape remains a helix before and after deformation, an analytical formula calculating the tape strain has been proposed[32, 33]. This formula has been adopted in many studies which apply tensile or torsional loads to CORC. Anvar et al. estimate the tape strain and the degradation of critical current during CORC tension, based on analytical and FEM models[34]. Wang et al. analyze the current-carrying properties of CORC with multi-layer tapes, and discuss the relationship between tape strain and interlaminar contact stress[33]. Besides tensile loads, many researchers also applied torsional loads to CORC. Ye et al. study the current degradation of CORC during torsion. In the CORC samples with two layers of tapes, they find some creases occurring in the tapes' gaps[35]. Yan et al. simulated the electro-mechanical behavior of CORC with multi-layer tapes, and provide some advice for the optimization of CORC[36]. Zheng et al. analytically calculated the strain of tape as it twists with the core. The effects of geometric parameters on the behavior of CORC are discussed, by combining the theoretical model of tape production, winding and CORC torsion[37]. Generally speaking, our study on HFRC in this paper can be based on the above researches. However, the deformation of a spiral tube (core of HFRC) is more complicated than that of a solid rod (core of traditional CORC). For example, the radial deformation of spiral tube can be influenced by its helical angle under tension or torsion, while the radial deformation of solid rod is relatively stable, attributed to the absence of "helical angle" in the design parameters of solid rod. As a result, some differences may arise in the tape strain of HFRC and CORC. Therefore, the deformation properties of spiral tube should be firstly studied, then the design of HFRC can be further optimized.

In this paper, we firstly introduce some formulas for calculating the strain of tape in a single-layer CORC. At the same time, a simplified theoretical model is established, which describes the radial deformation of spiral tube under tension or torsion. Then, the FEM model of CORC/HFRC is introduced. Based on the above analysis, relevant physical processes are simulated, and the reliability of FEM and theoretical model is verified against each other. Finally, the mechanical behavior of multi-layer HFRC under tension or torsion is studied according to the results of simulation. Considering the deformation characteristics of spiral tube, we also discuss the strain distribution of tapes in HFRC with different helical angle of spiral tube.

## 2. Simplified theoretical model

### 2.1 Tape strain of a single-layer CORC

In order to test the reliability of the FEM models in this paper, the tension and torsion processes of CORC with one layer of tape (single-layer CORC) are firstly simulated. Then, the tape strain along its centerline calculated by simulation is compared with relevant results calculated by theoretical formulas. In this section, these formulas are introduced based on some existing studies[33, 36-38].

In this paper, the rotation angle per unit length $\varphi$ is specified as positive when the vector of applied rotational displacement coincides with the outward normal of the cable end (Fig 2). For the convenience of analysis, a micro-segment is taken from the centerline of tape. After deformation, the projection of micro-segment in the core's cross-section corresponds to the central angle of $\Delta L_0 \varphi + \dfrac{\Delta L_0 \cot \alpha_0}{R_0}$ and the radius of $R_0(1+\varepsilon_r)$ (Fig 2). Hence, the length of this micro-segment before and after deformation $\Delta S_{tape}^0, \Delta S_{tape}$ can be expressed as:

$$\Delta S_{tape}^0 = \Delta L_0 \csc \alpha_0 \tag{1}$$

$$\Delta S_{tape} = \Delta L_0 \sqrt{(1+\varepsilon_t)^2 + (\varphi R_0 + \cot \alpha_0)^2 (1+\varepsilon_r)^2} \tag{2}$$

where $\varepsilon_t$ is the tensile strain applied to the whole cable, $\varphi$ is the rotation angle of cable per unit length along its axis, and $\varepsilon_r = (R-R_0)/R_0$ is the radial strain of core. Then, the tape strain along its centerline $\varepsilon_{tape}$ can be determined[33, 38]:

$$\varepsilon_{tape}(\varepsilon_t, \varphi, \varepsilon_r) = \frac{\Delta S_{tape} - \Delta S_{tape}^0}{\Delta S_{tape}^0} = \sin\alpha_0 \sqrt{(1+\varepsilon_t)^2 + (\varphi R_0 + \cot\alpha_0)^2 (1+\varepsilon_r)^2} - 1$$
$$\approx \varepsilon_t \sin^2\alpha_0 + \varphi R_0 \sin\alpha_0 \cos\alpha_0 + \varepsilon_r \cos^2\alpha_0 \quad (3)$$
$$(0 < \alpha_0 < \pi, \quad (1+\varepsilon_t)^2 \geq \scs^2\alpha_0 - (\varphi R_0 + \cot\alpha_0)^2 (1+\varepsilon_r)^2)$$

The part after "≈" in Eq. (3) originates from Taylor expansion [33]. It can be found that $\varepsilon_{tape}$ increases together with $\varepsilon_r$ if $\varepsilon_t$ and $\varphi$ remain constant. Therefore, the radial deformation of the core is considered to be a crucial factor affecting the tape strain, when the tensile or torsional deformation of cable has been known.

It should be noted that Eq. (3) is only applicable when $\varepsilon_{tape} \geq 0$. According to the equilibrium condition of tape, "$\varepsilon_{tape} < 0$" means that the interaction between the tape and core is traction stress. However, given their contact relationship, the core can only apply extrusion stress to the tape. To avoid this conflict, relevant variables need to satisfy: $(1+\varepsilon_t)^2 \geq \scs^2\alpha_0 - (\varphi R_0 + \cot\alpha_0)^2 (1+\varepsilon_r)^2$.

There are some differences in the radial deformation between CORC tension and torsion. When the CORC (solid core) is subjected to torsion, strain within the core mainly reflects in shear strain, and the radial strain $\varepsilon_r$ of core is close to zero under the condition of small deformation. However, when the CORC is subjected to axial tension, the radial deformation is relatively obvious due to Poisson effect. In our model, the core is considered to be a kind of elastic-linear-hardening material (copper), and it is in the state of uniaxial stress (ignoring the slight influence of tape) under tensile load. Therefore, the axial elastic and plastic strain of the core $\varepsilon_{core-elas}, \varepsilon_{core-plas}$ can be expressed as:

$$\varepsilon_{core-elas} = \frac{\sigma_y + E_{tan}\left(\varepsilon_t - \frac{\sigma_y}{E}\right)}{E} \quad (4)$$

$$\varepsilon_{core-plas} = \varepsilon - \varepsilon_{core-elas} \quad (5)$$

where $\sigma_y, E$ and $E_{tan}$ are respectively the yield strength, Young's modulus and tangent modulus of core. Then, the radial strain $\varepsilon_r$ of core can be obtained according to the theory of Plasticity:

$$\varepsilon_r = \frac{(R-R_0)}{R_0} = -\mu\varepsilon_{elas} - \frac{1}{2}\varepsilon_{plas} \tag{6}$$

where $\mu$ is Poisson's ratio of the core. By substituting Eq. (6) into Eq. (3), we can finally get the strain along the centerline of tape.

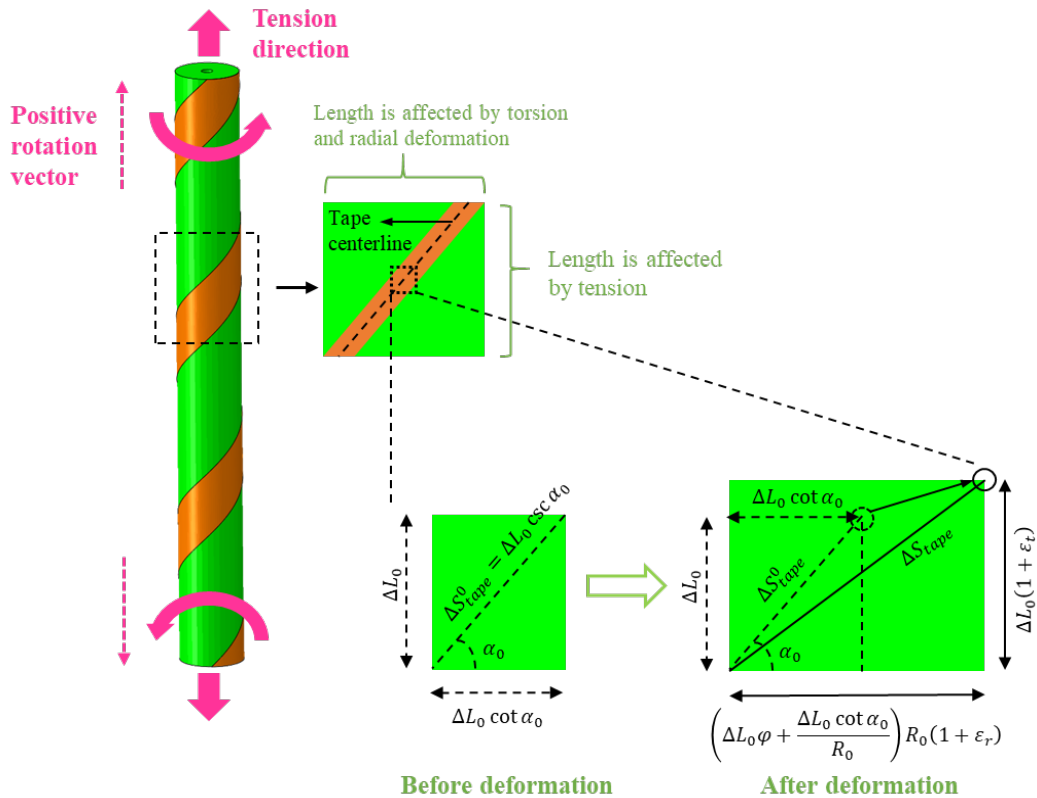

Fig. 2 Tape strain during the tension or torsion process of a single-layer CORC cable

## 2.2 Radial deformation of a single spiral tube

It can be easily found that the strain level of REBCO tapes in CORC or HFRC with multi layers of tapes can be approximately estimated by a formula similar to Eq. (3). Although the $R_0$ and $\varepsilon_r$ in this case should refer to the helical radius and radial strain of a tape itself (instead of the core), the radial deformation of core (solid rod or spiral tube) is also a crucial factor affecting the tape strain. This is because the radial deformation of the tapes in a certain layer is positively correlated with that of the core. In other words, an increase or decrease in the core' radius leads to a radial expansion or contraction of a tape in a certain layer. Correspondingly, this will lead to an increase or decrease in the tape strain, according to the relationships reflected in Eq. (3). As a basis for later research, a simplified analytical model describing the radial deformation of a single spiral tube (core of HFRC) is built in this subsection.

The ideal goal of the mechanical analysis of spiral tube is to build a theoretical model, which reflects the details of stress-strain distribution in each part of the tube. However, considering the special shape of a tube's cross-section (approximate sector perpendicular to the centerline of tube, Fig. 3(a)), some warpage may occur in the cross-section during deformation. This brings some difficult in the theoretical analysis. On the other hand, instead of more deformation details, only the radius of spiral tube under tension or torsion needs to be calculated in our analysis. Therefore, the configuration of spiral tube can be simplified to a spiral line in the section. Based on daily experience, a spiral tube or spiral line deforms more flexibly than a cylinder with the same dimensions. This is because the mechanical response of a helical structure is more of a rigid rotation and its line strain tends to be small. Hence, some extreme hypotheses can be proposed: (1) the tangential strain along the spiral line is zero; (2) the shape of the structure is maintained as a spiral line before and after deformation, accompanied by a change in the radius.

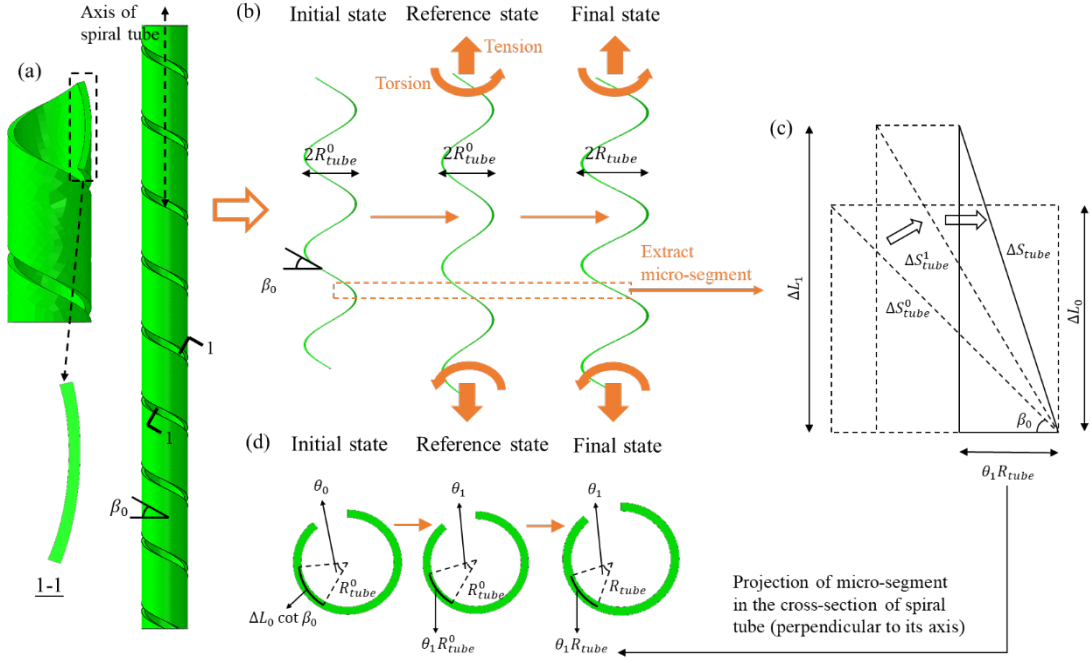

Fig. 3 Radial deformation of a single spiral tube. (a) Cross-section of the spiral tube perpendicular to its centerline, (b) simplification of the configuration of spiral tube, (c) deformation of a micro-segment of the spiral line, and (d) cross-section of the spiral tube perpendicular to its axis

Fig.3 shows the shape of the spiral line (simplification of spiral tube) under tensile and torsional displacement loads. In this paper, the helical direction of spiral line is uniformly left-handed and the helical angle is $\beta_0 \left( \beta_0 > 0 \right)$. As can be seen, a micro-segment is taken from the line. For analyzing convenience, the configuration of line is divided into three states: initial state, reference state and final state (Fig. 3(b)). Before deformation, the length of micro-segment is $\Delta S_{tube}^0$, while the projection of $\Delta S_{tube}^0$ in axial direction and in the cross-section of tube (perpendicular to its axis, Fig. 3(c)) are $\Delta L_0$ and $\Delta L_0 \cot \beta_0$ respectively. Hence, the central angle $\theta_0$ corresponding to $\Delta L_0 \cot \beta_0$ and the length of $\Delta S_{tube}^0$ can be presented as:

$$\theta_0 = \frac{\Delta L_0 \cot \beta_0}{R_{tube}^0} \tag{7}$$

$$\Delta S_{tube}^0 = \Delta L_0 \csc \beta_0 \tag{8}$$

where $R_{tube}^0$ is the radius of tube in initial state.

In reference state, the spiral line is considered to be subjected to tension or torsion without the variation of the radius. The length of micro-segment changes from $\Delta S_{tube}^0$ to $\Delta S_{tube}^1$. The projection of $\Delta S_{tube}^1$ in axial direction $\Delta L_1$ and the central angle corresponding to the projection of $\Delta S_{tube}^1$ in cross-section of tube (perpendicular to its axis, Fig. 3(c)) $\theta_1$ are:

$$\Delta L_1 = \Delta L_0 (1 + \varepsilon_t) \tag{9}$$

$$\theta_1 = \theta_0 - \varphi \Delta L_0 \tag{10}$$

where $\varphi$ is the rotation angle per unit length. Then, $\Delta S_{tube}^1$ can be expressed as:

$$\Delta S_{tube}^1 = \sqrt{(\Delta L_1)^2 + (\theta_1 R_{tube}^0)^2} \tag{11}$$

Considering the hypotheses that the tangential strain along the spiral line is zero, the configuration further changes from reference state to final state, meanwhile $\Delta S_{tube}^1$ should be turned into $\Delta S_{tube}$. The geometric relationship reveals that $\Delta S_{tube}$ and $\Delta S_{tube}^1$ have the same projected length ($\Delta L_1$) in axial direction. Similarly, their projections in the cross-section of spiral tube (perpendicular to its axis, Fig. 3(c)) correspond to the same central angle ($\theta_1$), but their radius change from $R_{tube}^0$ to $R_{tube}$. Subsequently, the length of $\Delta S_{tube}$ can be given as:

$$\Delta S_{tube} = \sqrt{(\Delta L_1)^2 + (\theta_1 R_{tube})^2} \tag{12}$$

The relationship between $\Delta S_{tube}$ and $\Delta S_{tube}^0$ can be listed according to the condition that the tangential strain along spiral line is zero:

$$\Delta S_{tube} - \Delta S_{tube}^0 = 0 \tag{13}$$

Combing Eq. (7)-(10) and Eq. (12)-(13), we can derive:

$$\sqrt{(\frac{\Delta L_0 \cot \beta_0}{R_{tube}^0} - \varphi \Delta L_0)^2 (R_{tube})^2 + \Delta L_0 (1+\varepsilon_t)^2} - \Delta L_0 \csc \beta_0 = 0 \tag{14}$$

$$(0<\beta_0<\frac{\pi}{2}, 0\leq \varepsilon_t < \csc \beta_0 -1, \varphi < \frac{\cot \beta_0}{R_{tube}^0})$$

From Eq. (14), the radius of tube after deformation $R_{tube}$ and the radial strain $\varepsilon_{tube-r}$ can be obtained:

$$R_{tube} = \frac{R_{tube}^0 \sqrt{(\csc \beta_0)^2 - (1+\varepsilon_t)^2}}{\cot \beta_0 - \varphi R_{tube}^0} \tag{15}$$

$$\varepsilon_{tube-r} = \frac{R_{tube}}{R_{tube}^0} - 1 = \frac{\sqrt{(\csc \beta_0)^2 - (1+\varepsilon_t)^2}}{\cot \beta_0 - \varphi R_{tube}^0} - 1 \tag{16}$$

It should be noted that Eq. (15) and (16) is only used to make an approximate estimate of the radial deformation, since the moment within the cross-section of tube (approximate sector perpendicular to the centerline of tube, Fig. 3(a)) may have an effect on the overall configuration.

## 3. FEM model

We use the commercial software Abaqus to complete our simulations. As shown in Fig. 4, the FEM model of HFRC is composed of a spiral tube core and several layers of REBCO tapes wound on the core, with three tapes in one layer. The outer and inner radius of the core are 2.8mm and 2.3mm, and its helical angle is about 27°. The width of the tape is 4mm and its winding angle is 45°/135°. The winding direction between adjacent layers is opposite. It is noted that the REBCO tape has a laminated structure, and its stacking order from inside to outside is substrate, buffer, superconducting layer, overlayer, and stabilizer[39-41]. In this paper, we build the FEM model of tape based on SCS4050[42, 43] produced by Superpower Inc. Because the buffer, superconducting layer and coverlayer are very thin, they are incorporated into the stabilizer and the material parameters of the stabilizer are given directly to this combined layer.

The thickness of each component layer is shown in Fig. 4(b), and the material parameters of spiral tube and tapes are listed in Table 1.

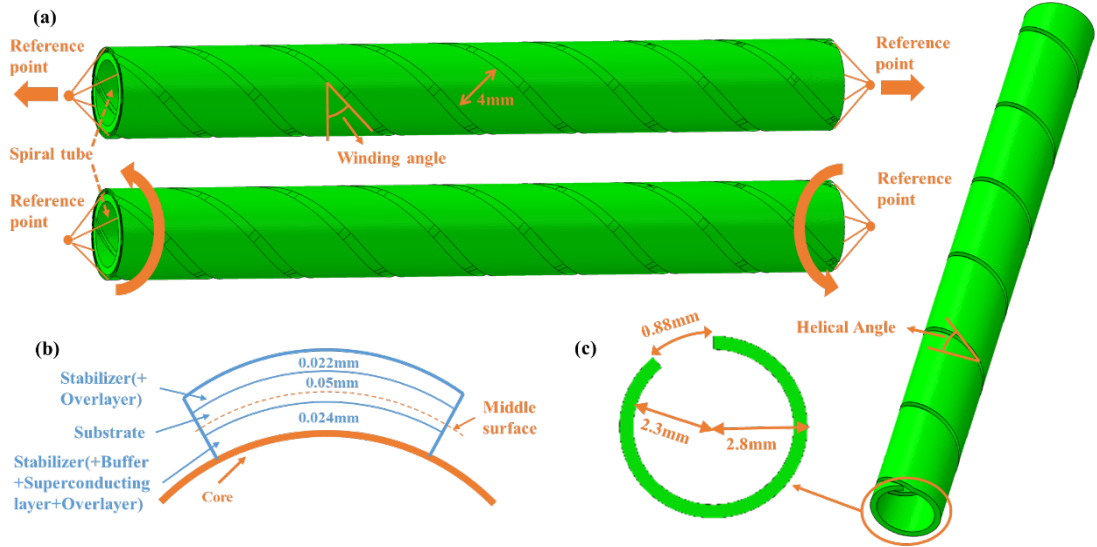

Fig. 4 FEM model of HFRC cable. (a) tension or torsion loads applied on HFRC, (b) geometry of REBCO tape and (c) geometry of spiral tube

Table 1 Material parameters of cable core and REBCO tape[44, 45]

| Material | Young's modulus (GPa) | Poisson's ratio | Yield strength (MPa) | Tangent modulus (GPa) | Density (kg×m$^{-3}$) |
|---|---|---|---|---|---|
| Copper (solid core/copper tape/stabilizer of REBCO tape) | 110 | 0.34 | 350 | 5 | 8300 |
| Hastelloy (substrate of REBCO tape) | 180 | 0.307 | 1225 | 7.5 | 8690 |
| Stainless Steel (spiral tube) | 193 | 0.3 | 840 | 10 | 8000 |

The friction coefficient between different layers of tapes (and the core) is 0.2. The core is meshed by solid element C3D8R, while the tape is meshed by shell element S4R. There are a total of 24 elements along the width of the tape. For applying external loads, the ends of the core and tapes are coupled with a reference point together (Fig. 4(a)). When the cable is tensioned, the reference point in each end of the cable translates in opposite directions along the cable axis. When the cable is twisted, these points rotate in opposite directions, and the rotation axis coincides with the cable axis. In addition to HFRC, we also simulate the tension and torsion processes of CORC, in order to compare their mechanical behavior. The FEM model of CORC is similar to that of HFRC, except that the core has been changed from a spiral tube to a round rod (solid core). To improve the calculation speed, the program is executed parallelly in Abaqus 2020, and the number of the processor and parallel domain is 12.

## 4. Model Validation

In this section, we respectively simulate the torsion and tension of a single-layer CORC cable and a single spiral tube in Abaqus. The reliability of single-layer CORC FEM model is verified by relevant results computed by the formula in subsection 2.1, giving a basis for later simulation of CORC/HFRC with multi-layer tapes. At the same time, the simulation results of a single spiral is used to validate the reasonableness of the theoretical analysis in subsection 2.2, so that the effects of the spiral tube on tape strain can be further discussed.

### 4.1 Torsion or tension of a single-layer CORC

When a single-layer CORC is under torsion only, the applied tensile strain $\varepsilon_t$ and radial strain $\varepsilon_r$ of core are close to zero. In this case, Eq. (3) degenerates to:

$$\varepsilon_{tape-torsion}(\varphi) = \sin\alpha_0 \sqrt{1+(\varphi R_0 + \cot\alpha_0)^2} - 1 \qquad (17)$$

Similarly, when the single-layer CORC is under tension only, the rotation angle per unit length $\varphi$ is close to zero. Therefore, Eq. (3) degenerates to:

$$\varepsilon_{tape-tension}(\varepsilon_t, \varepsilon_r) = \sin\alpha_0 \sqrt{(1+\varepsilon_t)^2 + \cot^2\alpha_0 (1+\varepsilon_r)^2} - 1 \qquad (18)$$

The variation of the tape strain along its centerline $\varepsilon_{tape}$ during torsion and tension are respectively shown in Fig 5(a) and Fig 5(b). The solid lines represent the formula estimation according to Eq. (17) and Eq. (18), while the dots represent the average values of tape strain in FEM simulations. For the accuracy and stability of strain distribution, the Dynamic-Implicit solver is used in Abaqus 2020 in this section.

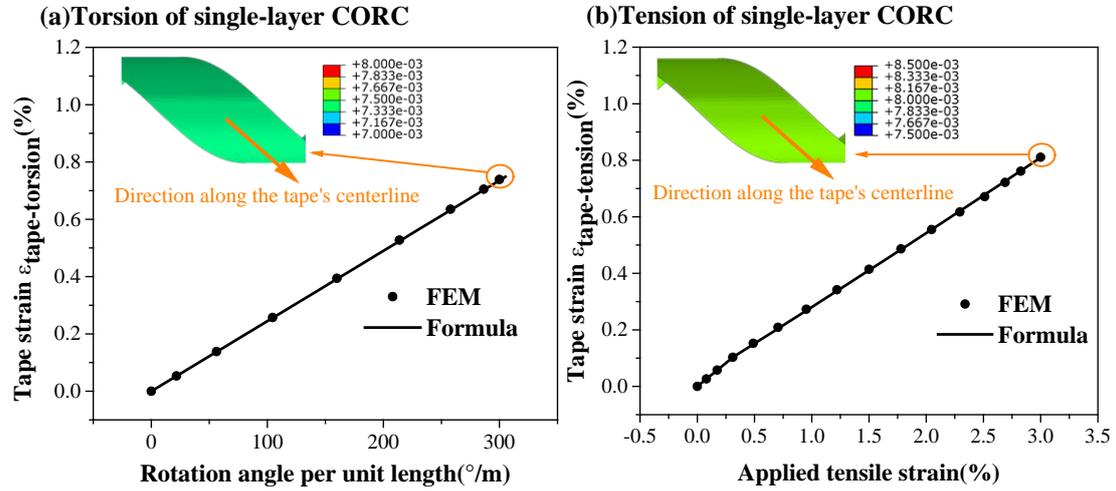

Fig. 5 The variation of tape strain in single-layer CORC during the process of (a) torsion and

(b) tension

From Fig 5, it can be seen that the tape train of single-layer CORC obtained from FEM is consistent with that from the formula estimation. Therefore, the reliability of the numerical model built in this paper is verified. Based on this, the behavior of multi-layer CORC and HFRC (spiral core) can be further discussed in Section 5.

**4.2 Torsion or tension of a single spiral tube**

When a single spiral tube is only subjected to torsion without tension, the applied tensile strain $\varepsilon_t$ in Eq. (15) is zero, and Eq. (15) degenerates to:

$$R_{tube-torsion} = \frac{R_{tube}^0 \cot\beta_0}{\cot\beta_0 - \varphi R_{tube}^0} \tag{19}$$

Likewise, when a single spiral tube is only subjected to tension without torsion, the rotation angle per unit length $\varphi$ is zero, and Eq. (15) degenerates to:

$$R_{tube-tension} = R_{tube}^0 \tan\beta_0 \sqrt{(\csc\beta_0)^2 - (1+\varepsilon_t)^2} \tag{20}$$

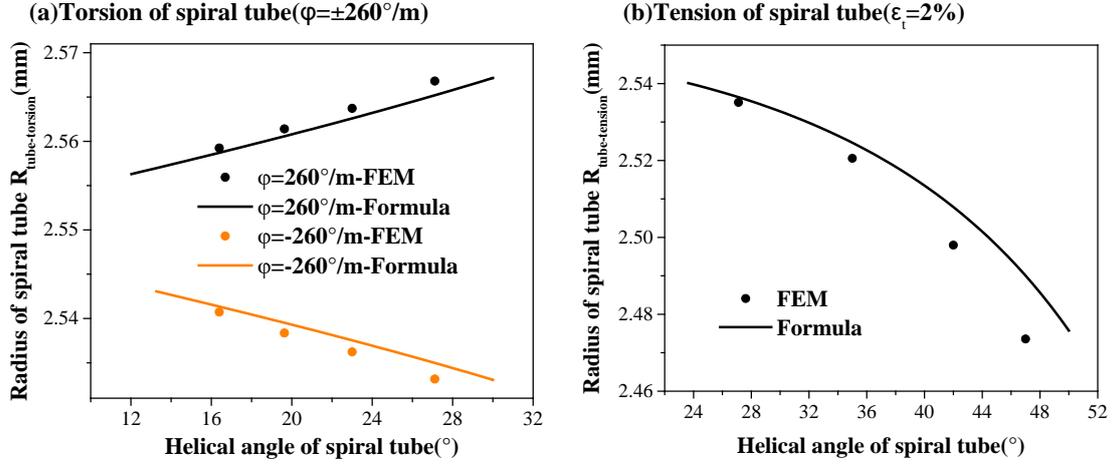

Fig. 6 The variation of radius of single spiral tube under external loads (a) torsion, (b) tension

In our simulation, torsional and tensile loads are applied to the spiral tube. The values of the radius after deformation $R_{tube}$ are extracted from simulation results and are compared with the values calculated by Eq. (19) and Eq. (20) in Fig 6. The efficiency of our theoretical model is validated by the similar trend of the FEM model. In the case of tension, $R_{tube}$ decreases with the increase of helical angle (Fig 6(c)). While in the case of torsion, the variation of $R_{tube}$ is related to the direction of torsional load. When the torsional direction is opposite to tube's helical direction ($\varphi = 260°/m > 0$), $R_{tube}$ is positively correlated with the helical angle (Fig 6(a)). When the torsional direction is consistent with the tube's helical direction ($\varphi = -260°/m < 0$), $R_{tube}$ is negatively correlated with the helical angle (Fig 6(a)). On the other hand, as stated in subsection 2.2, an increase or decrease in $R_{tube}$ will result in a larger or smaller tape strain. Therefore, the internal strain of tape in HFRC may be adjusted by changing the helical angle of core. This idea will be further discussed in next section.

In addition, it is easy to find from Fig 6. that there are some differences between the formula and FEM results. This should be attributed to the fact that simplifying the spiral tube to a spiral line ignores the deformation of its cross-section.

## 5. Results and discussion

In this section, the torsion and tension processes of multi-layer HFRC/CORC are simulated. For easier converge in calculation, the Dynamic-Explicit solver is used in Abaqus 2020. Based on simulation results, the properties of tape strain in HFRC (spiral core) are compared with those in CORC (solid core). Besides, the effects of the core's helical angle on tape strain in HFRC are also discussed. To demonstrate the variation of strain along the length of tape, we extract the strain values on the a section of the tape's centerline, which are shown in Fig 9(b, d), Fig 10(b), Fig 12(b), Fig 13(b) and Fig 15(b, d). The path of data extraction is shown in Fig 7. In addition, Fig 7 also indicates the direction of the train values we consider in this paper, which is parallel to the tape's centerline.

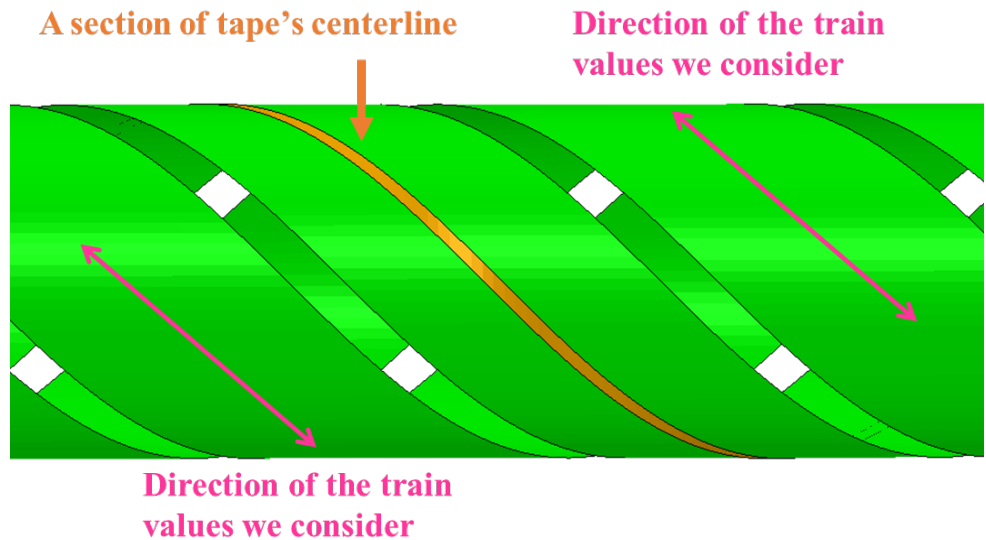

Fig. 7 The path for extracting train values and the direction of the tape we are concerned with

### 5.1 Torsion of HFRC or CORC

#### 5.1.1 Torsional direction is opposite to spiral tube's helical direction

In view of the structure of HFRC shown in Fig 1, the winding/helical direction of the tapes with even layer number (4th layer, 6th layer…) is the same as that of the core, while the winding/helical direction of the tapes with odd layer number (3th layer, 5th layer…) is contrary to that of the core. In this subsection, the direction of external torsional load is opposite to the core's helical direction (Fig 8(c)), which means the tapes with odd layer number are mainly subjected to stretching along their centerlines. To ensure that the outermost tapes does not detach from the cable surface, the total number of the tapes' layers needs to be odd. As shown in Fig 8, 5 layers of tapes are wound on the spiral tube. Since the inner two layers are copper tapes, we focus on the strain distribution of the outer three layers (REBCO tapes). The rotation angle per unit length $\varphi$ is 260°/m.

For solid core, the radius remains almost constant when the rotation angle is small. For spiral tube, radial expansion occurs during deformation and its radius increment will rise with the increase of its helical angle, according to our theoretical analysis (Eq. (19) and Fig 6(a)). On the other hand, Eq. (3) reveals that the tape strain will increase during the radial expansion of core (the increase of $\varepsilon_r$ and core's radius). Therefore, two conjectures about the tensile strain of the tapes with odd layer number arises: (1) The tape strain in HFRC (spiral core) is higher than that in traditional CORC (solid core); (2) As the helical angle of spiral tube becomes larger, the tape strain also increases. In Fig 9, the above conjectures are verified by simulation results.

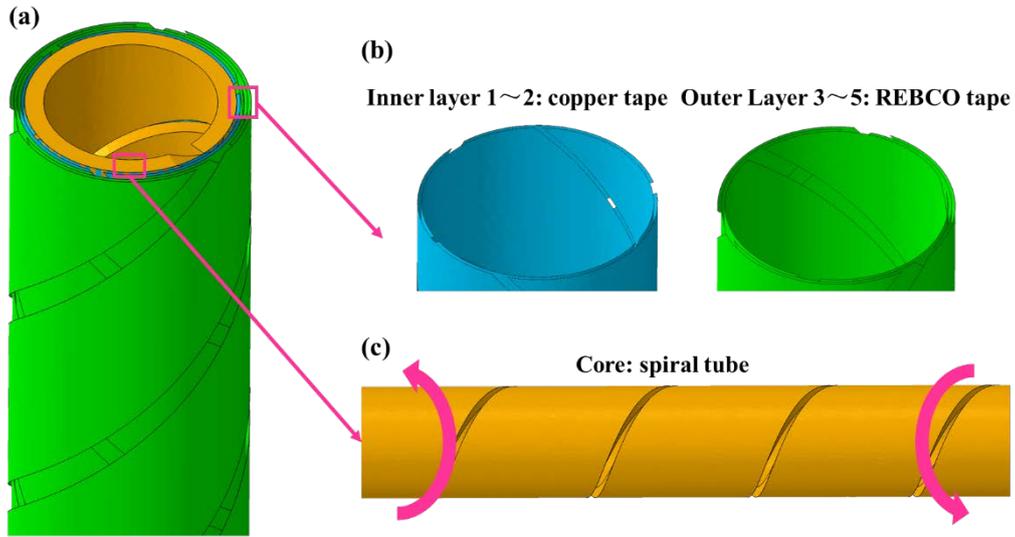

Fig. 8 The structure and loading way of HFRC: torsional direction is opposite to spiral tube's helical direction

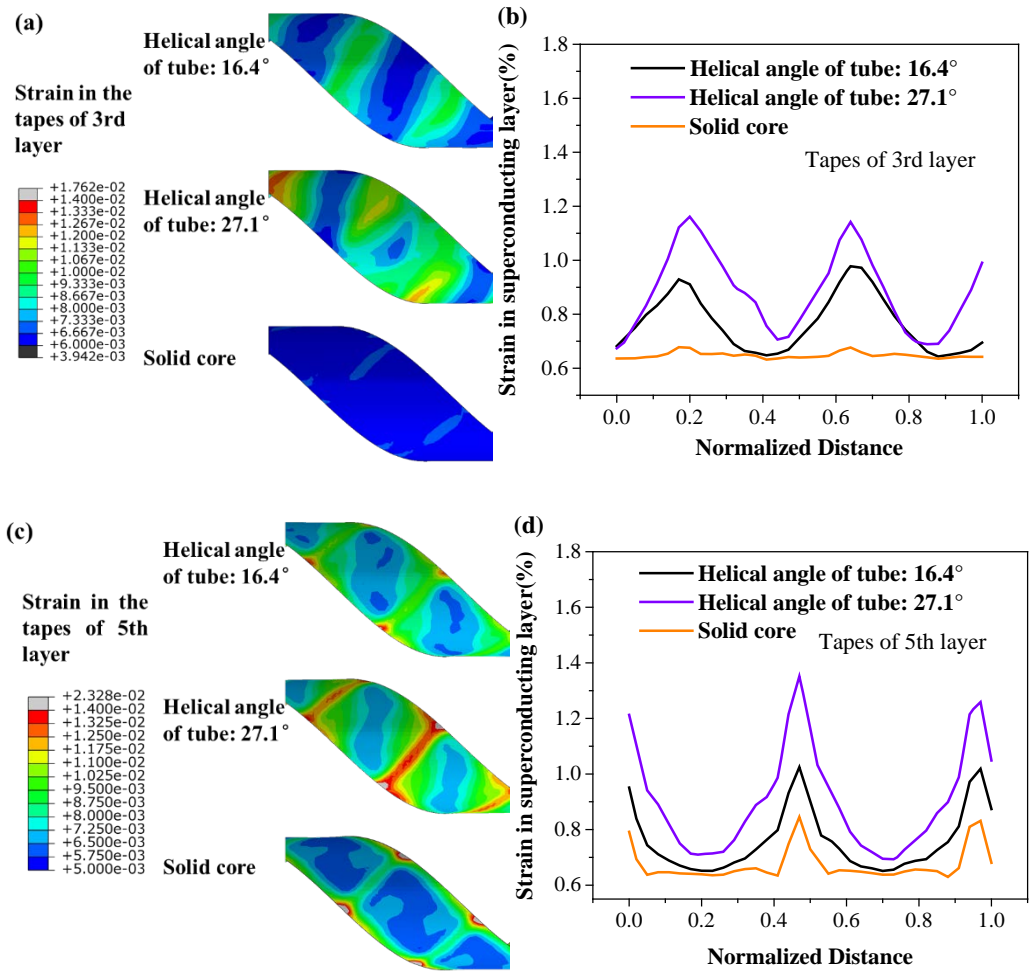

Fig. 9 Strain distribution in the tapes with odd layer number: torsional direction is opposite to spiral tube's helical direction. (a, b) Tapes of 3rd layer and (c, d) tapes of 5th layer

The strain distribution in the tapes with even layer number (4th layer) is displayed in Fig 10. In this case, the external torsional direction is opposite to the winding direction of tapes, so that the tapes are mainly subjected to compression along the centerlines. Under this condition, the tapes (4th layer) tend to deform unsteadily and some creases may be created between the gaps of the tapes of 5th layer. Similar phenomena have been observed in the experiments and simulations of traditional CORC (solid core) in [35, 36]. In this paper, we intend to investigate the characterization of such phenomena in HFRC (spiral tube). Here, the creases are presented in Fig 10(a, b). Considering that the superconducting layer is closer to the core than the middle surface of tape (Fig 4), strain in superconducting layer at creases is negative. As mentioned before, the spiral tube expands radially under torsion, which in turn limits the out-of-plane deformation of tapes. As a result, the degree of creasing and the strain in superconducting layer at creases is smaller in HFRC (spiral tube) compared with CORC (solid core).

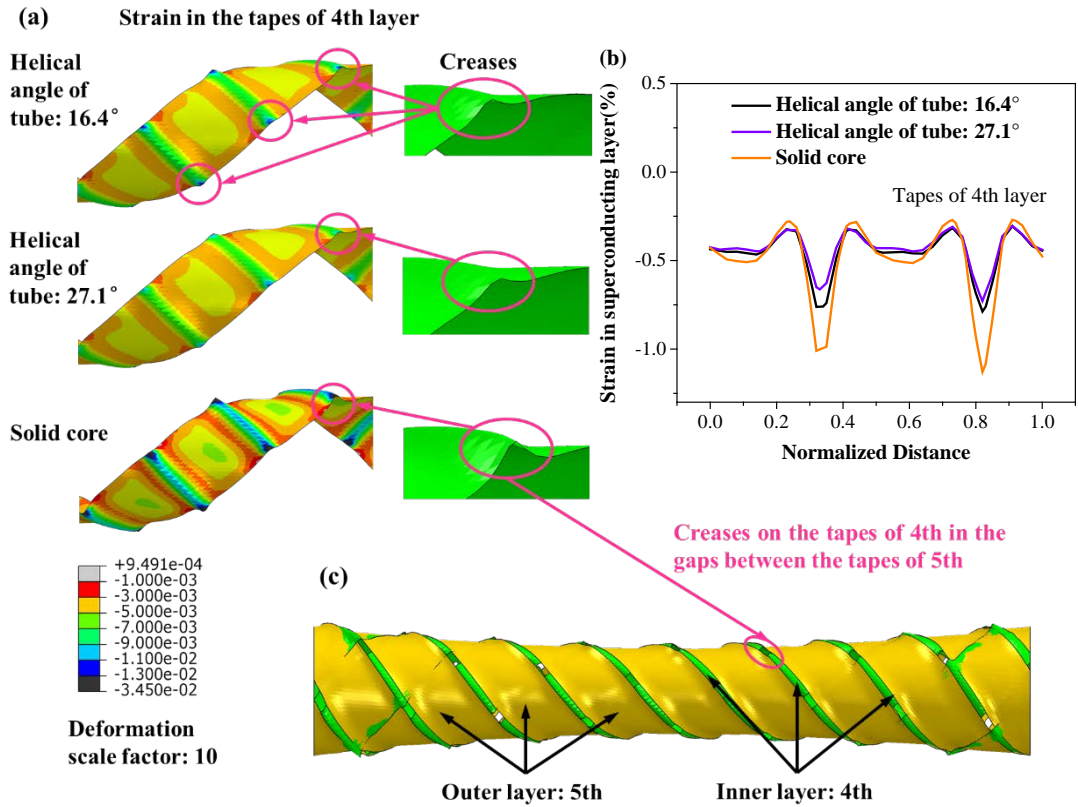

Fig. 10 Deformation and strain distribution in the tapes with even layer number : torsional direction is opposite to spiral tube's helical direction.

### 5.1.2 Torsional direction is consistent with spiral tube's helical direction

As mentioned in subsection 5.1.1, the winding/helical direction of the tapes with even layer number is the same as that of spiral tube. In this subsection, the direction of external torsional load is consistent with the core's helical direction (Fig 11(c)), which means the tapes with even layer number are mainly subjected to stretching along the centerlines. To ensure that the outermost tapes does not detach from the cable, the total number of tapes' layers needs to be even. As shown in Fig 11, 4 layers of tapes are wound on the spiral tube. Since the inner two layers are copper tapes, we focus on the strain distribution of outer two layers (REBCO tapes). The rotation angle per unit length $\varphi$ is -260°/m.

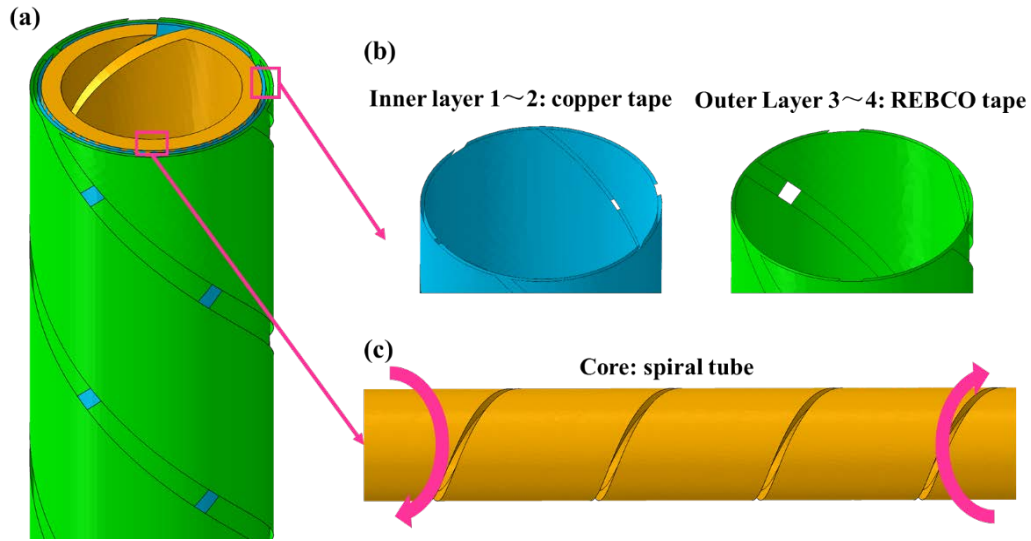

Fig. 11 The structure and loading way of HFRC: torsional direction is consistent with spiral tube's helical direction

The deformation of the tapes with odd layer number (3r layer) is displayed in Fig 12. In this case, the external torsional direction is opposite to the winding direction of tapes, so that they are mainly subjected to compression along centerlines. As mentioned in subsection 5.1.1, some creases may be created on these tapes (3rd layer). Considering the external torsional direction in this subsection, radial contraction will occur in the spiral tube during deformation, so its limitation on the out-of-plane deformation in tapes is reduced. Hence, the degree of creasing in HFRC (spiral tube) is larger than that in CORC (solid core).

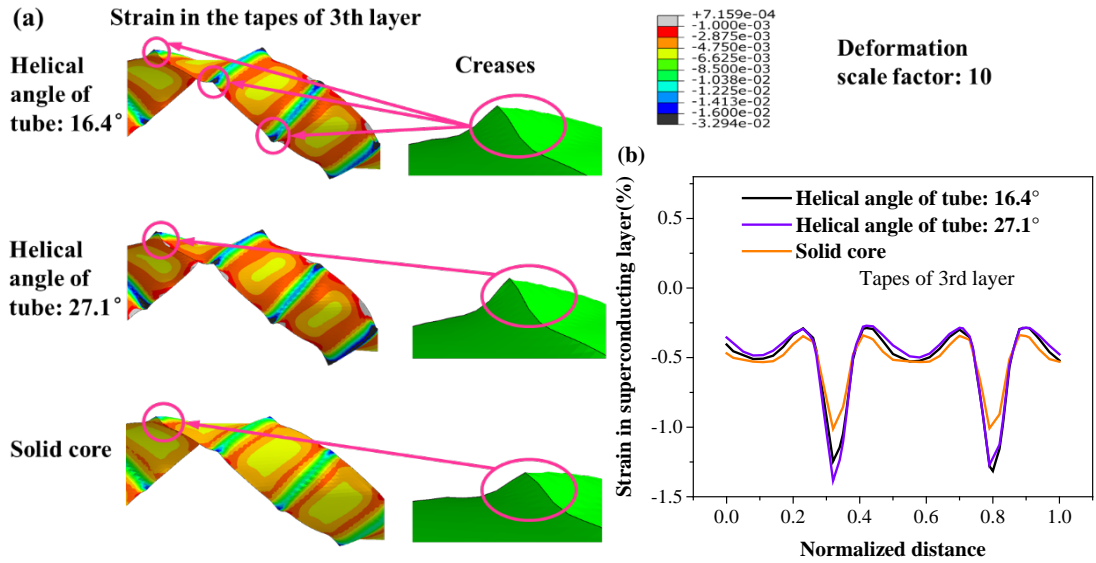

Fig. 12 Deformation and strain distribution of the tapes with odd layer number: torsional direction is consistent with spiral tube's helical direction.

The strain distribution in the tapes with even layer number (4th layer) is displayed in Fig 13. For solid core, the radius remains almost constant when the rotation angle per unit length $\varphi$ is small. For spiral tube, radial contraction occurs during deformation. According to Eq. (3), the tape strain is positively related to the radial strain of the core. Therefore, the tape strain is smaller in HFRC compared with CORC. In addition, it can be found in Fig 6(a) that the radius after deformation decreases with the increase of helical angle. This means that the tape strain can become smaller, when the helical angle is larger (Fig 13).

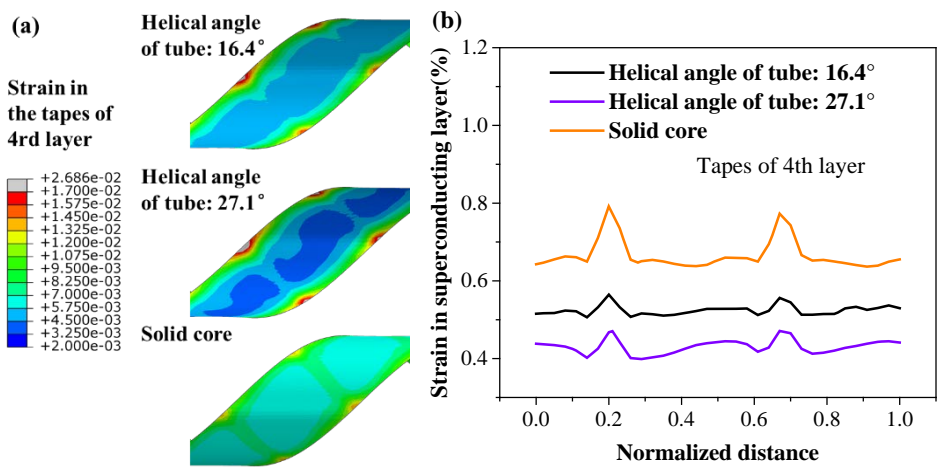

Fig. 13 Strain distribution in the tapes with even layer number: torsional direction is consistent with spiral tube's helical direction

## 5.2 Tension of HFRC or CORC

In the model of HFRC tension or CORC tension, four layers of tape are wound around the core. The inner two layers are copper tapes, while the outer two layers we are focused on are REBCO tapes (Fig 14). The applied tensile strain is 2%. In this case, tapes with either an odd or even layer number are subjected to stretching along their centerlines. Therefore, no crease is created in this case.

According to Fig 6(b), the radius of spiral tube after tension decreases with the increase of helical angle. At the same time, the tape strain is positively related to the radial strain of core according to Eq. (3). Thus, we conject that the tape strain becomes smaller for larger helical angle. The strain in a region of the tape obtained from FEM simulation is presented in Fig 15. These results verifie our conjecture. When the helical angle is 27.1°, the tape strain both in 3rd and 4th layer in HFRC is larger than that in CORC. When the helical angle increases to 42°, the tape strain in HFRC reduces to less than in CORC.

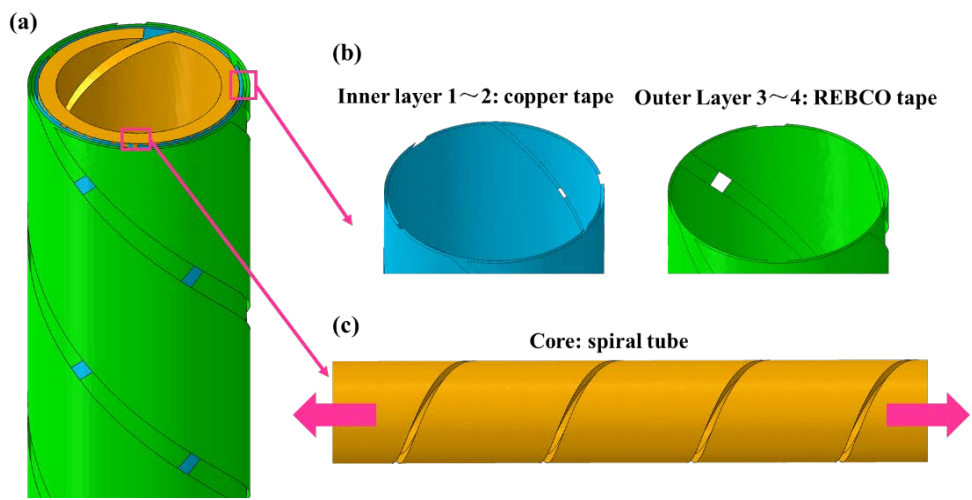

Fig. 14 The structure and loading way of HFRC: tension

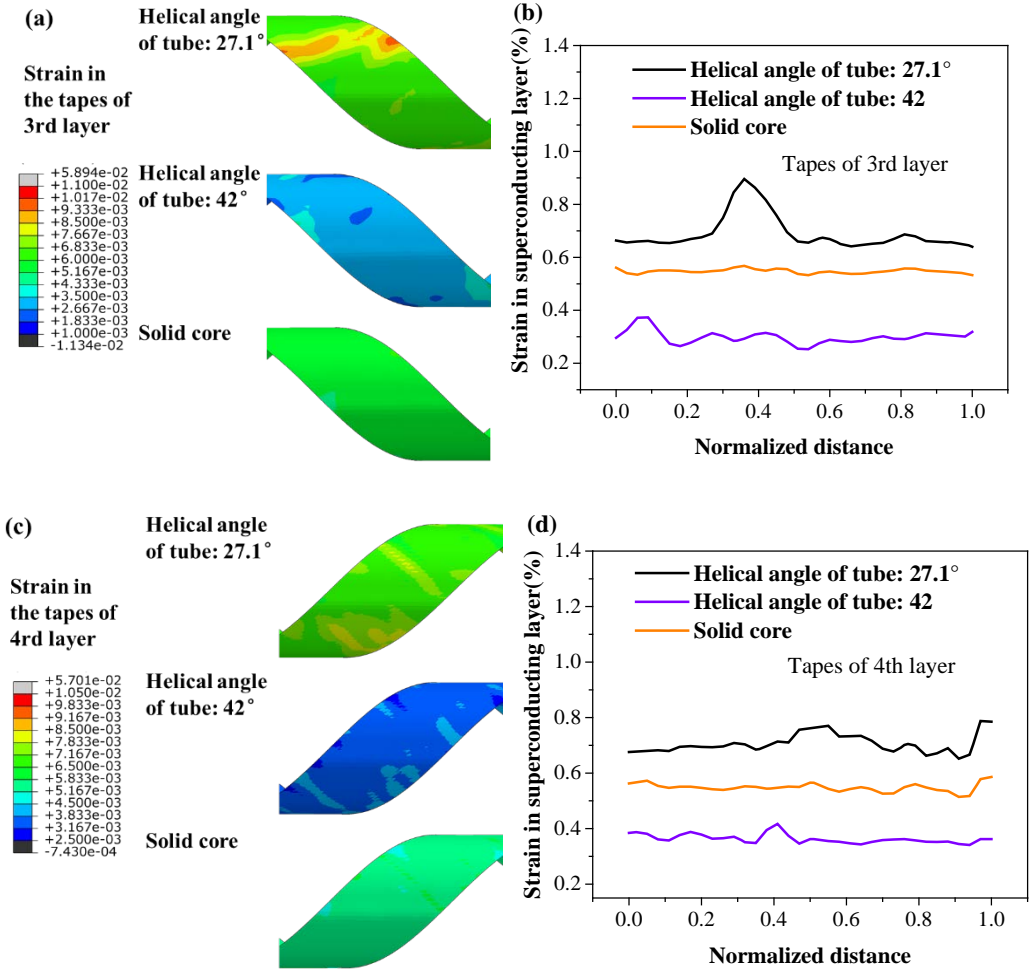

Fig. 15 Strain distribution in the tapes with tensile load. (a,b) Tapes of 3rd layer and (c,d) tapes of 4th layer

## 6. Conclusion

In this paper, we introduce the formulas for calculating the tape strain in CORC subjected to torsion or tension, referring to relevant research[33, 36-38]. The formulas reveal that the tape strain is positively related to the radial strain of cable core (or the radius of core after deformation). On the other hand, the radius of a single spiral tube (core of HFRC) after torsion or tension is calculated analytically. Relevant discussions indicate the radius after deformation is affected by the helical angle of tube. Considering the above, we think there is a relationship between the tape strain and the helical angle of tube.

According to the loading ways, the deformation properties of tapes in HFRC can be divided into three categories:

(1) Torsion (loading direction is opposite to spiral tube's helical direction): For the tapes with odd layer number, tensile strain in superconducting layer along the tape's centerline is raised in HFRC (spiral core) compared with CORC (solid core). Larger helical angle of core leads to higher tape strain. For the tapes with even layer number, the creases can be reduced by the radial expansion of core in HFRC.

(2) Torsion (loading direction is consistent with spiral tube's helical direction): For the tapes with odd layer number, larger creases occur in HFRC instead of CORC. For the tapes with even layer number, tensile strain is lower in HFRC compared with CORC, and larger helical angle results in lower tensile strain.

(3) Tension: The tensile strain of tape is negatively correlated with the helical angle of core. With different helical angle of core, the tape strain in HFRC is higher or lower than that in CORC.

In summary, which is more advantageous, HFRC cable or traditional CORC cable, depends on the loading ways, the tapes' position and the core's helical angle. In practice, the optimized design parameters should be chosen according to application requirements.

**Acknowledgements**

The authors acknowledge the supports from the National Natural Science Foundation of China (Nos. U2241267, 12172155 and 11932008), National Key Research and Development Program of China (No. 2023YFA1607304) and Major Scientific and Technological Special Project of Gansu Province (23ZDKA0009).

# References

[1] HAHN, S., KIM, K., KIM, K., HU, X., PAINTER, T., DIXON, I., KIM, S., BHATTARAI, K. R., NOGUCHI, S., JAROSZYNSKI, J., and LARBALESTIER, D. C. 45.5-tesla direct-current magnetic field generated with a high-temperature superconducting magnet. *Nature*, 570(7762), 496-499 (2019)

[2] MIYAZAKI, H., IWAI, S., OTANI, Y., TAKAHASHI, M., TOSAKA, T., TASAKI, K., NOMURA, S., KURUSU, T., UEDA, H., NOGUCHI, S., ISHIYAMA, A., URAYAMA, S., and FUKUYAMA, H. Design of a conduction-cooled 9.4 T REBCO magnet for whole-body MRI systems. *Superconductor Science and Technology*, 29(10), 104001 (2016)

[3] TANG, Y., LIU, D., LI, D., YONG, H., and ZHOU, Y. A modified model to estimate the screening current-induced magnetic field of a REBCO magnet. *Superconductor Science and Technology*, 35(4), 045013 (2022)

[4] TAKAYASU, M., CHIESA, L., BROMBERG, L., and MINERVINI, J. V. HTS twisted stacked-tape cable conductor. *Superconductor Science and Technology*, 25(1), 014011 (2011)

[5] MUZZI, L., DE MARZI, G., DI ZENOBIO, A., and DELLA CORTE, A. Cable-in-conduit conductors: lessons from the recent past for future developments with low and high temperature superconductors. *Superconductor Science and Technology*, 28(5), 053001 (2015)

[6] GOLDACKER, W., FRANK, A., KUDYMOW, A., HELLER, R., KLING, A., TERZIEVA, S., and SCHMIDT, C. Improvement of superconducting properties in ROEBEL assembled coated conductors (RACC). *IEEE Transactions on Applied Superconductivity*, 19(3), 3098-3101 (2009)

[7] WANG, Y., BAASANSUREN, S., XUE, C., and HASEGAWA, T. Development of a Quasi-Isotropic Strand Stacked by 2G Wires. *IEEE Transactions on Applied Superconductivity*, 26(4), 4804406 (2016)

[8] PENG, X., and YONG, H. Structural analysis of REBCO coated conductors and quasi-isotropic strands under bending using continuum shell element. *Cryogenics*, 133, 103701 (2023)

[9] VAN DER LAAN, D. C. $YBa_2Cu_3O_{7-\delta}$ coated conductor cabling for low ac-loss and high-field magnet applications. *Superconductor Science and Technology*, 22(6), 065013 (2009)

[10] VAN DER LAAN, D. C., WEISS, J. D., and MCRAE, D. M. Status of CORC® cables and wires for use in high-field magnets and power systems a decade after their introduction. *Superconductor Science and Technology*, 32(3), 033001 (2019)

[11] FAN, L., SONG, P., XIAO, M., SHAO, L., FENG, F., GUAN, M., and QU, T. Design and testing of a prototype Canted-Cosine-Theta HTS dipole magnet using CORC cable. *IEEE Transactions on Applied Superconductivity*, 34(3), 4600605 (2024)

[12] WU, J., LIU, D., ZHANG, X., and YONG, H. Mechanical Response of Conductor on Round Core (CORC) Cables Under Electromagnetic Force. *Acta Mechanica Solida Sinica*, 36(3), 418-427 (2023)

[13] LI, Y., MU, N., TANG, S., ZHANG, Z., ZHOU, J., YONG, H., and ZHANG, X. Deformation and crack prediction of CORC cable induced by Poisson effect: Theoretical modeling and experimental validation. *Engineering Fracture Mechanics*, 292, 109625 (2023)

[14] LI, X., TANG, Y., XU, Y., and REN, L. Analytical analysis of hollow CORC cable under thermo-mechanical loads. *Superconductivity*, 5, 100037 (2023)

[15] RIES, R., GöMöRY, F., MOŠAŤ, M., KUJOVIČ, T., HINTZE, C., and GIL, P. Effect of off-axis bending on microstructural and transport properties of coated conductor tape. *Superconductor

*Science Technology*, 36, 014006 (2023)

[16] WANG, S., YONG, H., and ZHOU, Y. Numerical calculations of high temperature superconductors with the J-A formulation. *Superconductor Science and Technology*, 36(11), 115020 (2023)

[17] SHAN, S., WANG, S., YONG, H., and ZHOU, Y. Numerical simulations of electromagnetic behavior in CORC cable based on a modified H-φ formulation. *Superconductor Science and Technology*, 36(5), 055006 (2023)

[18] SHAN, S., YONG, H., and ZHOU, Y. Mechanical behavior of multi-layer CORC cable in high external field with 3D numerical model. *Physica C: Superconductivity and its Applications*, 620, 1354501 (2024)

[19] WANG, S., YONG, H., and ZHOU, Y. Calculations of the AC losses in superconducting cables and coils: Neumann boundary conditions of the T–A formulation. *Superconductor Science and Technology*, 35(6), 065013 (2022)

[20] JIN, H., WU, Q., XIAO, G., ZHOU, C., LIU, H., TAN, Y., LIU, F., and QIN, J. Bending performance analysis on YBCO cable with high flexibility. *Superconductivity*, 7, 100054 (2023)

[21] LI, M., ZHENG, J., KHODZHIBAGIYAN, H., MA, T., HUANG, X., NOVIKOV, M., and CHENG, Y. Engineering Design of Forced-Flow Cooling HTS Cable for SMES System With High Current Capacity. *IEEE Transactions on Applied Superconductivity*, 34(5), 5700105 (2024)

[22] SHI, Y., MA, T., DAI, S., JIN, H., and QIN, J. Bending performance of the CORC cable with flexible interlocked stainless steel former. *Superconductor Science and Technology*, 36(11), 115011 (2023)

[23] LIU, H., JIN, H., XIAO, G., ZHOU, C., LIU, X., LIU, F., LIU, H., and QIN, J. Experimental Study on the Cabling Performance of REBCO Cable With Different Twist Pitches. *IEEE Transactions on Applied Superconductivity*, 34(5), 4803705 (2024)

[24] GUO, Z., QIN, J., LUBKEMANN, R., WANG, K., JIN, H., XIAO, G., LI, J., ZHOU, C., and NIJHUIS, A. AC loss and contact resistance in highly flexible rebco cable for fusion applications. *Superconductivity*, 2, 100013 (2022)

[25] XIAO, G., JIN, H., ZHOU, C., MA, H., WANG, D., LIU, F., LIU, H., NIJHUIS, A., and DEVRED, A. Performance of highly flexible sub-cable for REBCO Cable-In-Conduit conductor at 5.8 T applied field. *Superconductivity*, 3, 100023 (2022)

[26] SHI, Y., MA, T., DAI, S., and LIU, W. Experimental research on the influence of HTS tape width and former type on the transverse compression performance of CORC cables. *Physica C: Superconductivity and its Applications*, 619, 1354463 (2024)

[27] LI, Y., WEI, L., and ZHANG, X. Measurement of Nonuniform Strain Distribution in CORC Cable Due to Bending Process by a Segmentation-Aided Stereo Digital Image Correlation. *Experimental Mechanics*, 63(5), 813-822 (2023)

[28] SHI, Y. Y., DAI, S. T., MA, T., LIU, W. X., JIN, H., and QIN, J. G. Analysis on the transverse compression performance of the CORC cable. *Superconductor Science and Technology*, 35(12), 125005 (2022)

[29] VAN DER LAAN, D. C., RADCLIFF, K., ANVAR, V. A., WANG, K. Y., NIJHUIS, A., and WEISS, J. D. High-temperature superconducting CORC® wires with record-breaking axial tensile strain tolerance present a breakthrough for high-field magnets. *Superconductor Science and Technology*, 34(10), 10LT01 (2021)

[30] TANG, S., PENG, X., and YONG, H. Numerical simulation of the mechanical behavior of

superconducting tape in conductor on round core cable using the cohesive zone model. *Applied Mathematics and Mechanics*, 44(9), 1511-1532 (2023)

[31] MA, J., and GAO, Y. Mixed-mode fracture analysis of CORC cables with double-edge cracks under tension and torsion. *Engineering Fracture Mechanics*, 307, 110277 (2024)

[32] LANTEIGNE, J. Theoretical Estimation of the Response of Helically Armored Cables to Tension, Torsion, and Bending. *Journal of Applied Mechanics*, 52(2), 423-432 (1985)

[33] WANG, K. Y., GAO, Y. W., ANVAR, V. A., RADCLIFF, K., WEISS, J. D., VAN DER LAAN, D. C., ZHOU, Y. H., and NIJHUIS, A. Prediction of strain, inter-layer interaction and critical current in CORC® wires under axial strain by T-A modeling. *Superconductor Science and Technology*, 35(10), 105012 (2022)

[34] ANVAR, V. A., WANG, K. Y., WEISS, J. D., RADCLIFF, K., VAN DER LAAN, D. C., HOSSAIN, M. S. A., and NIJHUIS, A. Enhanced critical axial tensile strain limit of CORC® wires: FEM and analytical modeling. *Superconductor Science and Technology*, 35(5), 055002 (2022)

[35] YE, H. S., ZHOU, X., YUAN, Y. C., HU, R., YANG, J. S., JIN, Z. Q., ZHAO, Y., and SHENG, J. Study on torsion behavior of superconducting conductor on round core cable. *IEEE Transactions on Applied Superconductivity*, 32, 4801005 (2022)

[36] YAN, J. T., WANG, K. Y., and GAO, Y. W. Numerical analysis of the mechanical and electrical properties of CORC cables under torsional loading. *Cryogenics*, 129, 103624 (2023)

[37] ZHENG, K., and GOU, X. Analytical Investigation of Axial Strain of the YBCO Tape on CORC Winding Cables Under Twisting Deformation. *IEEE Transactions on Applied Superconductivity*, 33(9), 8401307 (2023)

[38] PENG, Y., and GOU, X. Strain analysis of the superconductor REBCO tape of CORC cables under winding, bending and twist deformations. *Acta Mechanica Sinica*, 40, 723245 (2023)

[39] ZHAO, Y., ZHU, J. M., JIANG, G. Y., CHEN, C. S., WU, W., ZHANG, Z. W., CHEN, S. K., HONG, Y. M., HONG, Z. Y., JIN, Z. J., and YAMADA, Y. Progress in fabrication of second generation high temperature superconducting tape at Shanghai Superconductor Technology. *Superconductor Science and Technology*, 32(4), 044004 (2019)

[40] SUGANO, M., MACHIYA, S., SATO, M., KOGANEZAWA, T., SHIKIMACHI, K., HIRANO, N., and NAGAYA, S. Bending strain analysis considering a shift of the neutral axis for YBCO coated conductors with and without a Cu stabilizing layer. *Superconductor Science and Technology*, 24(7), 075019 (2011)

[41] GAO, P., ZHANG, Y., WANG, X., and ZHOU, Y. Interface properties and failures of REBCO coated conductor tapes: Research progress and challenges. *Superconductivity*, 8, 100068 (2023)

[42] WANG, K. Y., TA, W. R., and GAO, Y. W. The winding mechanical behavior of conductor on round core cables. *Physica C: Superconductivity and its Applications*, 553, 65-71 (2018)

[43] ANVAR, V. A., ILIN, K., YAGOTINTSEV, K. A., MONACHAN, B., ASHOK, K. B., KORTMAN, B. A., PELLEN, B., HAUGAN, T. J., WEISS, J. D., VAN DER LAAN, D. C., THOMAS, R. J., PRAKASH, M. J., HOSSAIN, M. S. A., and NIJHUIS, A. Bending of CORC® cables and wires: finite element parametric study and experimental validation. *Superconductor Science and Technology*, 31(11), 115006 (2018)

[44] DAS, I., SAHOO, V., and RAO, V. V. Structural Analysis of 2G HTS Tapes under Different Loading Conditions for HTS Power Cable using Finite Element Modeling. *Physica C: Superconductivity and its Applications*, 618, 1353771 (2024)

[45] NIU, M. D., XIA, J., YONG, H. D., and ZHOU, Y. H. Quench characteristics and mechanical

responses during quench propagation in rare earth barium copper oxide pancake coils. *Applied mathematics and mechanics (English edition)*, 42(2), 235-250 (2021)